\documentstyle[prl,aps]{revtex}

\newcommand{\vol}[1]{{\bf #1}}

\newcommand{\ttitle}[1]{{\it #1}}
\newcommand{\pretitle}[1]{{\em #1},}
\newcommand{\prenumber}[1]{Report No. #1}

\newcommand{\tselea}[1]{\label{#1}}
\newcommand{\tseleq}[1]{\label{#1}} 
\newcommand{\tbib}[1]{\bibitem{#1}} 
\newcommand{\tref}[1]{(\ref{#1})}
\newcommand{\tcite}[1]{\cite{#1}} 

\newcommand{\tnote}[1]{}
\newcommand{\tpre}[1]{}

\newcommand{\href}[2]{#2}
\newcommand{\eprint}[1]{{#1}}

\renewcommand{\href}[2]{{#2}{}}
\renewcommand{\eprint}[1]{\href{http://xxx.soton.ac.uk/abs/#1}{{\tt #1}}}
\renewcommand{\tpre}[1]{#1}

\newcommand{\half}{\frac{1}{2}}
\newcommand{\bea}{\begin{eqnarray}}
\newcommand{\eea}{\end{eqnarray}}
\newcommand{\beq}{\begin{equation}}
\newcommand{\eeq}{\end{equation}}
\newcommand{\beqn}{\[}
\newcommand{\eeqn}{\]}

\newcommand{\Ez}{E}
\newcommand{\Ed}{E'}
\newcommand{\Edd}{E''}
\newcommand{\taui}{\tau_i}
\newcommand{\tauz}{\tau}
\newcommand{\taud}{\tau'}
\newcommand{\taudd}{\tau''}
\newcommand{\nz}{n}
\newcommand{\nd}{n'}
\newcommand{\ndd}{n''}

\newcommand{\Bret}{B_{\rm ret}}
\newcommand{\Seff}{S_{\rm eff}}
\newcommand{\Sefft}{S_{\rm eff}^{(2)}}
\newcommand{\intdtk}{\int \frac{d^3\veck}{(2\pi)^3}  \;}
\newcommand{\intdtp}{\int \frac{d^3\vecp}{(2\pi)^3}  \;}
\newcommand{\intdtx}{\int d^3\vecx  \;}
\newcommand{\intdfk}{\int \frac{d^4k}{(2\pi)^4}  \;}

\newcommand{\intdfx}{\int d^4x \;}

\newcommand{\intdtauz}{\int_{\taui}^{\taui-i\beta} d\tauz \;}
\newcommand{\intdtaud}{\int_{\taui}^{\taui-i\beta} d\taud \;}
\newcommand{\veck}{\vec{k}}
\newcommand{\vecp}{\vec{p}}
\newcommand{\vecx}{\vec{x}}
\newcommand{\vecnab}{\vec{\nabla}}

\newcommand{\calB}{{\cal B}}
\newcommand{\calL}{{\cal L}}

\newcommand{\bbZ}{{\cal Z}}
\newcommand{\Real}{{\cal R}e}

\newcommand{\ebom}{e^{\beta \omega}}

\newcommand{\etdtom}{e^{-2i\tau\omega}}

\begin{document}

\draft

\preprint{\parbox{6cm}{{\tt Imperial/TP/97-98/21} 
\\ \eprint{hep-ph/9808382} } }

\title{Derivative expansions of Euclidean thermal effective
actions}

\author{T.S.Evans\thanks{email: \href{mailto:T.Evans@ic.ac.uk}{\tt T.Evans@ic.ac.uk}, 
\tpre{WWW: \href{http://euclid.tp.ph.ic.ac.uk/links/time}{\tt http://euclid.tp.ph.ic.ac.uk/\symbol{126}time} }}
}

\address{\href{http://euclid.tp.ph.ic.ac.uk/}{Theoretical Physics}, 
Blackett Laboratory, Imperial College,
Prince Consort Road, London SW7 2BZ,  U.K. }

\date{22nd February 1998, revised 16th July 1998}

\maketitle

\begin{abstract}
I compute the derivative expansion of an effective action at finite
temperature using the imaginary time approach.  I show that it is a
well behaved expansion giving a unique seriers contrary to previous
results.  This disparity is shown to originate in the choice of
thermal Green functions used in the calculations.
\end{abstract}

\pacs{PACS: 11.10.Wx, 11.10.Lm\tpre{. Preprint {\tt
Imperial/TP/97-98/21}, {\tt hep-ph/9808382}}}


Effective actions provide a powerful tool for analysing complex
systems.  They enable the complicated effects 
of different sectors of the theory to be incorporated in a compact 
way into a description of the one sector of interest.  
For example, in thermal field theory 
the effective action encodes the effects that the quantum and thermal 
fluctuations of several fields have on the specific field under 
consideration.
However, as effective actions of quantum fields are often very complicated, 
it is common to consider the low energy/momentum behaviour of one field by 
expanding the effective action in terms of the derivatives of
that field.
This is a well
trodden path in zero temperature quantum field theory 
\tcite{Co,Fr,DGMP}.\tnote{Coleman, equation (3.16) of ``Secret
Symmetries \ldots'', p.133}  

It is therefore surprising that similar calculations of derivative
expansions of thermal effective actions give strange results 
\tcite{AT,Fu2,TSEze,TSEwpg,EEV,GH,DH,AL}.   It has been  shown in
the literature that series is not unique as it depends on  the order
in which one performs the time and space derivative expansions.  In
particular the lowest order term, which is usually identified as the
effective potential or free energy \tcite{Co,Fr,DGMP}, varies
significantly.   The importance of this fact has been stressed in
the condensed matter literature, for instance \tcite{AT,AL}.  In
relativistic particle physics, such derivative expansions are also
frequently used but at finite temperature it is common to address
only the spatial derivatives \tcite{MTW}.  However, many practical
problems are dynamical ones, such as tunnelling rates, or the slow
roll of fields in inflation, so temporal derivatives are required.  

In this letter I show how to calculate a well behaved derivative
expansion of an effective action using the Euclidean approach to
thermal field theory.  The key idea is that such expansions describe
field configurations which vary {\em slowly} in time and space and
hence are not thermal equilibrium configurations.  By explicit
calculation, I will show that the retarded thermal Green functions
used in previous analyses are {\em not} relevant to this problem. 
These Green functions describe how the system responds to a {\em
sudden} impulse as linear response theory shows.  Important
physical processes, especially Landau damping, do contribute to my
unique derivative expansion whereas they caused the problems in the
literature where retarded Green functions were incorrectly used.

Recently, the problem was clearly demonstrated by Das and Hott \tcite{DH}
in a simple relativistic model, putting earlier comments 
\tcite{AT,Fu2,TSEze,TSEwpg,EEV,GH}\tnote{The end of the introduction
of \tcite{Fu} and less direct comments in \tcite{Fu2},  p.42 of
\tcite{TSEth}, after equation (4) of \tcite{TSEze} but nothing in
\tcite{TSEzm}, opening paragraph of \tcite{TSEwpg}.  \tcite{Vo} does
not mention effective actions, but does \tcite{AVBD} - No? 
\tcite{We2} does not mention effective actions.}  
into context. 
Following \tcite{DH} I use a model with two scalar fields 
\beqn
\calL [\phi,\eta] = \half \eta \Delta^{-1}{\eta}  - \half g \phi
\eta^2 + \calL_0 [\phi] 
\eeqn 
where $\Delta$ is
the $\eta$ field propagator  and $\calL_0 [\phi]$ is the free
Lagrangian for $\phi$ with arbitrary dispersion relation and source.
Treating $\phi$ as the background field, whose effective action I
wish to calculate, and integrating out the $\eta$ field fluctuations
gives a generating functional of the form 
\bea 
Z & = & \int D\phi \;
\exp \{i \int d^4x \calL_0[\phi] + i\Seff'[\phi] \} 
\nonumber \\
\Seff'[\phi] &=& \frac{i}{2} {\rm Tr} \left\{ \ln \left[ 1 - g 
\Delta(x,x') \phi(x')\right] \right\} 
\nonumber 
\eea
Four-vectors are always Minkowskii, e.g.\ $k^\mu = (E,\veck), k^2  =
E^2-\veck^2$, with $E\in {\Real}$ for physical values.  

The only non-trivial contribution to the
effective action comes from the non-local logarithmic term $\Seff'$
which I expand as a series in $g\phi$, 
$\Seff' = \sum_{p=1}^{\infty} \Seff^{(p)}$.  
The non-locality is first clearly visible in the $O(g^2\phi^2)$ term, $\Sefft$, 
\bea
\Sefft &=& \frac{-ig^2}{4}\intdfx \intdfx\!' \; \phi(x) \Delta(x,x') 
\phi(x') \Delta(x',x) 
\tselea{Seff20}
\eea
so we will focus on this term c.f.\ \tcite{DH}. 

The derivative expansion
is an attempt to represent such non-local terms as a series of 
local terms.  Thus $\phi(x')$ of \tref{Seff20} is expressed in terms
of $\phi(x)$ through a Taylor series, which we may write as
\bea
\phi(x') &=& 
\left. \exp \{ \lambda [(x'-x).\frac{\partial}{\partial{x''}}] \} \phi(x'')
\right|_{x''=x, \lambda=1}
\tselea{dexp1}
\eea
In \tref{dexp1} the formal counting parameter $\lambda$ 
allows us to generate the terms of the derivative expansion
as a $\lambda$ power series. This gives 
\bea
\Sefft 
&=&
-\half
\left.
\intdfx  \phi(x) B(-i\lambda\partial_\mu ) \phi(x)
\right|_{\lambda=1}
\tselea{dexp2}
\\
-i B(p) &=& \frac{(-ig)^2}{2}\intdfk i\Delta (k) i\Delta(k+p)
\nonumber 
\eea
Thus, all the information needed about the contribution of the
$\eta$ field fluctuations to the derivative expansion of $\Sefft$ 
comes from a small four-momentum expansion of the bubble diagram
$B(p)$.   The formal parameter $p^\mu = -i\lambda \partial^\mu$ 
is not in general small
so such a derivative expansion of the bubble diagram $B$ can only
have any meaning if $\phi(p)$ dies off sufficiently fast as $p^\mu$
moves away from zero, i.e.\ the expansion is valid only for {\em
small} deviations from space-time independent  $\phi$ field
configurations. At zero temperature with real Minkowskii or
Euclidean times and energies, this works perfectly
\tcite{Co,Fr,DGMP}.  

Now let us turn to finite temperature and use the Euclidean time
approach \tcite{lB}.  At first sight the arguments which led from
\tref{Seff20} to \tref{dexp2}  work equally well in thermal field
theory provided one changes the time integrations to run from
$\tau_i$ to $\tau_i-i\beta$, uses scalar fields which are periodic
over this range, and alters the propagators appropriately.

However, the precise meaning of $B$ is not clear in the thermal
context because the same diagram represents many different Green
functions with distinct values \tcite{BM,TSEnpt}.  Existing analyses
of the effective action in this language
\tcite{AT,Fu2,TSEze,TSEwpg,DH} (but not {\tcite{GH}) {\em assume}
without proof that retarded functions should be used.  This is the
most natural for Euclidean thermal field theory \tcite{BM,TSEnpt}
and the effective potential at finite temperature is most easily
linked to an expansion in terms of retarded diagrams \tcite{TSEwpg}.
 For later comparison, let us see what this gives for $\Sefft$.  In
\tref{Seff20} we set $B=\Bret$, the retarded bubble diagram, which
can be calculated with any thermal field formalism to give  
\bea
\Bret (p_\mu=(E,\vecp);\beta) &=& -\frac{g^2}{2}\intdtk
\sum_{s_0,s_1 =\pm 1} \frac{s_0s_1}{4 \omega \Omega} 
\frac{1}{E+s_0\omega + s_1\Omega}  \frac{(e^{\beta (s_0 \omega + s_1
\Omega)} - 1 ) }{ (e^{\beta s_0 \omega} -1)(e^{\beta s_1 \Omega}
-1)} 
\tselea{bres1} 
\eea 
where $\omega = \omega(\veck)$ and $\Omega = \omega(\veck+\vecp)$
and we have assumed $\Delta(E,\veck) = (E^2-\omega^2)^{-1}$. 

The problem with this thermal retarded diagram is that it does not
have a unique value in the $p^\mu \rightarrow 0$ limit
\tcite{AT,Fu2,TSEze,TSEwpg,EEV,GH,DH,AL,We2}.  In \tref{bres1}, it
comes from a divergence in the denominator when $E+s_0\omega + s_1
\Omega \rightarrow 0$ as $p^\mu \rightarrow 0 $ for the terms $s_0 =
- s_1$ when $\beta < \infty$.  A more familiar form for
the retarded bubble is obtained if we now assume a relativistic
dispersion relation and then perform the angular integration, but
this still has the same  problems in the limit of  $p^\mu
\rightarrow 0$. This strange behaviour of the retarded bubble
diagram at zero momentum is {\em not} due to any technical
calculational problems \tcite{TSEwpg,We2}.  From a physical point of
view the origin of this branch point is well known and arises from
the physical processes known as Landau damping \tcite{lB}.  These
processes have particles from the heat bath involved in the {\em in}
or {\em out} states and thus are only present at non-zero
temperature.

Physically, using $\Bret$ implies that the behaviour of fields
described by an effective action is {\em very} sensitive to the way I
approach the zero momentum point.  It implies that {\em almost
constant} fields, which vary only on scales much, much bigger than
any physical scale in the problem, have a significantly different
behaviour when they vary faster in space than time as compared to the
opposite case.
  
It is therefore worth performing the calculation more carefully,
focusing on the temporal aspect as this changes most radically at
finite temperature.  Starting from \tref{Seff20} I find  
\bea
\Sefft &=& \frac{-ig^2}{4}
\intdtauz \intdtp \phi(\tauz,\vecp) .
\nonumber \\
&&
\left. \intdtaud \intdtk  \Delta(\tauz,\taud;\veck) 
\phi(\taud,-\vecp) \Delta(\taud,\tauz;\veck+\lambda \vecp) 
\right|_{\lambda=1}
\tselea{Seff2t}
\eea
The propagators are for the $\eta$ field in thermal
equilibrium.  For illustrative purposes we again work with a
relativistic field for which
the most convenient representation here is
\beqn
\Delta (\tauz,\taud;\veck) 
=\frac{i}{\beta} \left. \sum_{n\in \bbZ}
\sum_{s=\pm 1} 
e^{-iE(\tauz-\taud)}
\frac{s}{2\omega} 
\frac{1}{E -s \omega (\veck)} \right|_{E=2\pi i n / \beta} 
\eeqn

The key point comes in the need to
expand the second field in \tref{Seff2t} as
 in \tref{dexp1}
\beq
\phi(\taud,-\vecp) 
=
\left. \exp \{ \lambda (\taud-\tauz) \frac{\partial}{\partial {\tau''}} \}   
\phi(\taudd,-\vecp)   \right|_{\taudd=\tauz, \lambda=1}
\tseleq{expand2}
\eeq
Performing the $\taud$ integration in $\Sefft$ \tref{Seff2t} 
gives
\bea
I &=& \intdtaud e^{i(\Edd+\Ez-\Ed)\taud} e^{-i(\Edd + \Ez-\Ed)\tauz}
= \frac{(e^{(\Edd + \Ez-\Ed)\beta} - 1)}{i(\Edd+\Ez-\Ed)}
e^{i(\Edd + \Ez-\Ed)(\tau_i-\tauz)} 
\tselea{I1}
\eea
where $\Edd \equiv -i \lambda (\partial /\partial \taudd)$ coming
from \tref{expand2}.  When we use the result $I$ in \tref{Seff2t}, 
$\Ez$ and $\Ed$ come from the two propagators and so  are both
identical to Matsubara frequencies $2 \pi i n / \beta, n\in \bbZ$. 
{\em If and only if} $\Edd$ is  also limited to Matsubara
frequencies does $I$  then reduce to the usual energy conserving 
delta function $I = - i \beta \delta_{\ndd,\nd-\nz}$. From
\tref{expand2} it is clear that taking $\Edd \equiv -i \lambda
(\partial/\partial \taudd )$ identically equal to
one of the Matsubara frequencies is equivalent to saying that the
background field $\phi$ must be periodic in imaginary time.  

However, I intend to truncate the derivative expansion of the $\phi$
field.  This is a disaster in the compact time coordinate  because
the truncated Taylor series of a periodic function is not
necessarily periodic!   A perturbation of interest to a periodic
background field $\phi(\taudd)$ may be small and periodic itself, 
but keeping only a finite numbers of terms in the Taylor series \tref{expand2}
leaves us with a non periodic  perturbation.  Thus I am forced to
consider energies of the background $\phi$ field,
$\Edd$ in \tref{I1}, which are merely in the neighbourhood of zero
energy and are not a pure Matsubara frequency, so that $\exp \{
\beta \Edd \} \neq 1$.  The derivative
expansion is good for studying {\em small, but arbitrary} deviations from an
constant background.  This includes non-periodic, i.e.\ non-equilibrium, perturbations
where for a short time the heat bath will remain in equilibrium.

To put the point more forcefully, existing analyses assume $\exp \{
\beta \Edd \} = 1$ and are therefore  only correct if they impose
equilibrium rigorously.  Consequently they are correct only for
static configurations and  expanding such results about zero energy
is invalid.  One exception is \tcite{GH} where it is also argued
that the limitation of $\Edd$ to Matsubara frequencies makes no
sense.  However, in \tcite{GH} no rigorous algorithm for the choice
of the analytic continuation in energy away from discrete $\Edd$ is
provided c.f.\ \tcite{BM,TSEnpt}, and Weldon \tcite{We2} has shown
the use of ``Feynman parameterisation'' in \tcite{GH} is incorrect. 

With non-periodic perturbations allowed, one can show that  the
calculation is no longer expected to be independent of the initial
time $\tau_i$, which is the time at which the temperature and
density matrix were defined \tcite{AG,TSEinprep}.  Taking great care
 over factors such as $\exp \{ \beta E\}$ and over the analytic
properties of integrands at large energies when energy sums over
$\Ez, \Ed$ are performed, contour integration techniques \tcite{lB}
give 
\bea 
{\Sefft} 
&=& 
-\frac{1}{2} \intdtauz \intdtp  \left.
\phi(\tauz,\vecp)  \calB ( -i\lambda\frac{\partial}{\partial
\taudd},\lambda\vecp;m;\tauz - \taui)  \phi(\taudd, -\vecp)
\right|_{\taudd=\tauz, \lambda=1} 
\tselea{Seff2r} 
\\ 
\calB  (\Edd,\vecp;m;\tauz - \taui)  
&=& 
\frac{g^2}{2} \intdtk
\sum_{s_0,s_1 = \pm 1}  \frac{ s_0 s_1}{4 \omega \Omega} 
\frac{1}{(\Edd + s_0 \omega + s_1 \Omega)} \frac{1}{(e^{\beta s_0
\omega} -1)(e^{\beta s_1 \Omega}-1)} . 
\nonumber \\ 
&& 
\left\{ 
(e^{\beta(\Edd + s_0 \omega + s_1 \Omega)} - 1) +
(e^{-i(\tauz-\tau_i)(\Edd + s_0 \omega + s_1 \Omega)} -1 ).
(e^{\beta\Edd} -1 ) e^{\beta (s_0 \omega_0 + s_1 \Omega)}  \right\}
\nonumber 
\eea 
Note that the earlier problems with the
derivative  expansion in \tref{bres1} stemmed from the  $\Edd + s_0
\omega + s_1 \Omega$ denominator, which in  the limit of static and
homogeneous fields ($\Edd \equiv -i \lambda(\partial
/\partial\taudd) = 0$, $\vecp=0$) is divergent.  The same
denominator is present in \tref{Seff2r} but now the numerators are
zero in this limit.  Thus the space and time derivative expansion of
this expression is well defined.   It also means that there are no
branch points at $\Edd^2 = \vecp^2$ or $\Edd^2 = 4m^2+\vecp^2$ though 
the latter can never be seen in a small $\Edd,\vecp$ expansion.

Equation \tref{Seff2r} shows that $\calB = \Bret$ of \tref{bres1} 
only if I set $\exp \{ \beta \Edd \} =1$, i.e.\ limit myself to
exactly periodic $\phi$ field configurations.  This again emphasises
the fact that the use of the retarded bubble diagram in the
derivative expansion for effective actions is valid only for static
configurations.

In order to make the spatial derivative expansion, I need the form
of the dispersion relation of the $\eta$ field.  Taking this to be
the usual relativistic form, $\omega(\veck)^2 = |\veck|^2
+ m^2$, it is straight forward to generate the expansion.  
Up to first order I have
\bea
{\Sefft}' &=& \frac{g^2}{32\pi^2} \intdtauz \intdtx 
\; \phi(\tauz, \vecx) \left[
\gamma_{0,0}  -i\beta \gamma_{1,0}(\tau-\tau_i)
\frac{\partial}{\partial \tau} + 
O( \frac{\partial^2}{\partial \tau^2}, \vecnab^2) 
\right] 
 \phi(\tauz,\vecx) 
\tselea{Sefftexp}
\\
\gamma_{0,0}  &=& 
\int_0^\infty dk \; \frac{-k^2}{\omega^3} \left[
\frac{\ebom +1}{\ebom -1}
+ 2\beta \omega
\frac{\ebom }{(\ebom -1)^2} \right] \tpre{= \frac{k^2}{\omega}
\frac{d}{d\omega} \left( \frac{1}{\omega} \coth (\half\beta \omega ) \right) }
\nonumber \\
\gamma_{1,0} (\tau)&=& \half \gamma_{0,0} 
+\int_0^\infty dk \; \left[ \frac{2k^2}{\omega^2} 
\frac{\ebom }{(\ebom -1)^2} (-i\tau)
+\frac{k^2}{2\omega^3} 
\frac{(\etdtom -1 )(1+ e^{2i(\tau - i \beta)\omega}) }{ (\ebom -1)^2} 
\right]
\nonumber
\eea
This first order expansion illustrates several features of the
general result \tref{Seff2r}.  Galliean invariance
tells us that there can be no terms odd in $\vecnab$, and no
such terms are present in \tref{Seff2r} or \tref{Sefftexp}.  
Terms with no time derivatives correspond to
pure static perturbations in the background fields.  These are
trivially periodic and hence \tcite{AG,TSEinprep} must be independent of
$\tauz-\taui$.  $\gamma_{0,0}$ and the explicit form \tref{Seff2r}
confirm this.  Interestingly, there is a term linear in the time derivative.  This
unfamiliar term is a Lorentz scalar at finite temperature because of the
existence of an extra four-vector, $u^\mu$, the four velocity of the
observer with respect to the heat bath.  I have
implicitly chosen to measure quantities with respect to the heat
bath rest frame so that $u^\mu = (1, \vec{0})^\mu$ and 
the relevant Lorentz scalars are $\partial_\mu \partial^\mu$ and
$\partial /\partial\tau \equiv \partial_\mu . u^\mu$.

Most importantly, the lowest order term is the contribution to the
effective potential and $\gamma_{0,0}$ is indeed found to be
precisely that obtained in standard calculations of the effective
potential at finite temperature, be they real- or imaginary-time
calculations.  Only now with this analysis and its well behaved
derivative expansion can we justify the use such a free energy term 
as a potential in an evolution equation for quantum fields in a heat bath,
something which has often been assumed at in the
past.

Contrary to previous suggestions, this result shows that Landau
damping does not destroy the derivative expansion.   Indeed, Landau
damping terms are present in the above calculation as they are
the $s_1=-s_2$ terms with $\omega-\Omega$ factors in \tref{Seff2r}. 

Further confirmation that this approach to derivative expansions of
thermal effective actions is correct comes from a real-time
calculation \tcite{AG} and from one using a general path-ordered
formalism \tcite{TSEreg,TSEinprep} calculation, both of which give the same
result as this Euclidean calculation.  Likewise, the generalization
to other types of field is straight forward \tcite{TSEinprep}.  

Finally, this analysis of thermal effective actions is not in
conflict with linear response \tcite{lB}, where retarded Green
functions are relevant.  In linear response one considers
perturbations which are space-time delta functions.  This is the
exact opposite of the nearly constant field configurations 
described by the derivative expansion of approximation to thermal
effective actions. Thus we should not be surprised when we find
Green functions like $\calB$ of \tref{Seff2r} and not retarded Green
functions control the dynamics of slowly varying fields.

I would like to thank I.Aitchison, M.Asprouli, V.Galan Gonzalez,
R.Rivers and D.Steer for useful discussions.  This work was
supported by the Royal Society, and by the European Commission under
their Human Capital and Mobility programme, contract number {\tt
CHRX-CT94-0423}.



\begin{references}

\tbib{Co} S.Coleman, \ttitle{Aspects of Symmetry} (Cambridge
University Press, Cambridge, 1985\tpre{, ISBN {\tt 0 521 31827 0}}).

\tbib{Fr} C.M.Fraser, Z.Phys.C \vol{28} 101 (1985).

\tbib{DGMP} A.Dobado, A.G\'{o}mez-Nicola, A.L.Maroto, J.R.Pel\'{a}ez, 
\ttitle{Effective Lagrangians for the Standard Model} 
(Springer-Verlag, Berlin, 1997\tpre{, ISBN 3-540-62570-4}).

\tbib{AT} E.Abrahams and T.Tsuneto, Phys.Rev.\ \vol{152} 416 (1966).

\tbib{Fu2} Y.Fujimoto, in \ttitle{Quantum Field Theory}, edited by F.Mancini
(Elsevier Science Publishers, Amsterdam, 1986), p.435.

\tbib{TSEze} T.S. Evans, Z.Phys.C \vol{41} (1988) 333.

\tbib{TSEwpg} T.S.Evans, Can.J.Phys. \vol{71} 241 (1993).

\tbib{EEV} P.Elmfors, K.Enqvist, and I.Vilja, Phys.Rev.Lett.
\vol{71} 480 (1993)\tpre{ (\eprint{hep-ph/9303269})}.

\tbib{GH} P.S.Gribosky and B.R.Holstein, Z.Phys.C \vol{47} 214 (1990).

\tbib{DH} A.Das and M.Hott, Phys.Rev.D \vol{50} 6655 (1994).

\tbib{AL} I.J.R.Aitchison and D.J.Lee, 
Phys.Rev.B \vol{56} 8303 (1997).

\tbib{MTW} I.Moss, D.Toms and A.Wight, 
Phys.Rev.D \vol{46} 1671 (1992).

\tbib{lB} M.LeBellac, \ttitle{Thermal Field Theory}
(Cambridge University Press, Cambridge, 1996\tpre{, ISBN {\tt 0 521 46040 9}}).

\tbib{BM} G.Baym and N.D.Mermin, 
J.Math.Phys.\ \vol{2} 232 (1961).

\tbib{TSEnpt} T.S.Evans, Nucl. Phys. \vol{B374} 340 (1992).

\tbib{We2} H.A.Weldon, Phys.Rev.D \vol{47} (1993) 594.

\tbib{AG} M.Asprouli and V. Galan Gonzalez, 
\pretitle{Derivative expansions of real-time thermal effective actions}
Imperial College, \tpre{Imperial/TP/97-98/026,} 
\prenumber{\eprint{hep-ph/9802319}}.

\tbib{TSEreg} T.S.Evans, \pretitle{The Unique Derivative Expansion for Thermal Effective
Actions} \tpre{talk given at the
\href{http://theory.uwinnipeg.ca/tft98.html}{5th International Workshop on Thermal Field
Theory},
Regensburg, Germany, 10-14 August, 1998. Original overheads
available at
\href{http://euclid.tp.ph.ic.ac.uk/links/time/papers/reg/regtalk.ps}{http://euclid.tp.ph.ic.ac.uk/\symbol{126}time/papers/reg/regtalk.ps}
,} \tpre{Imperial/TP/97-98/69,} \prenumber{\eprint{hep-ph/98mmnnn}}.

\tbib{TSEinprep} T.S.Evans, {\em in preparation.}

\end{references}
\end{document}